# Dynamic nuclear polarization and nuclear magnetic resonance in the simplest pseudospin quantum Hall ferromagnet

H. W. Liu,[1,2,*] K. F. Yang,[1] T. D. Mishima,[3] M. B. Santos,[3] and Y. Hirayama[1,4,†]

[1]*ERATO Nuclear Spin Electronics Project, Sendai, Miyagi 980-8578, Japan*

[2]*State Key Lab of Superhard Materials and Institute of Atomic and Molecular Physics, Jilin University, Changchun 130012, P. R. China*

[3]*Homer L. Dodge Department of Physics and Astronomy, University of Oklahoma, Norman, OK 73019-2061, USA*

[4]*Department of Physics, Tohoku University, Sendai, Miyagi 980-8578, Japan*

[*]liuhw@ncspin.jst.go.jp

[†]hirayama@m.tohoku.ac.jp

We present dynamic nuclear polarization (DNP) in the simplest pseudospin quantum Hall ferromagnet (QHF) of an InSb two-dimensional electron gas with a large $g$ factor using tilted magnetic fields. The DNP-induced amplitude change of a resistance spike of the QHF at large current enables observation of the resistively detected nuclear magnetic resonance of the high nuclear spin isotope $^{115}$In with nine quadrupole splittings. Our results demonstrate the importance of domain structures in the DNP process. The nuclear spin relaxation time $T_1$ in this QHF was relatively short (~ 120 s), and almost temperature independent.

Dynamic nuclear polarization (DNP), resulting from the transfer of spin polarization from electrons to nuclei, has recently attracted considerable attention because of its significance in the study of spin dynamics and spin-based information processing.[1] A two-dimensional electron gas (2DEG) at a semiconductor heterointerface is promising for this study. Although the coupling of nuclear spins to electrons in the single 2DEG is weak, resistively or optically detected nuclear magnetic resonance (RD NMR or ODNMR) has proven to be a highly sensitive probe of DNP.[2,3] Recent experiments have explored ingenious DNP methods for the coherent control of nuclear spins in a 2DEG[4,5] and for the discovery of fascinating 2-D electronic features (e.g., collective spin texture, low-frequency spin fluctuations, and quantum phase dynamics) in the integer or fractional quantum Hall (IQH or FQH) regime.[6-12] However, in the past 20 years, all of these DNP studies have been applied only to the GaAs 2DEG (each nucleus has nuclear spin $I = 3/2$), mainly because of the small $g$ factor (~ -0.44) and high mobility at low temperatures.

In this work, we present a novel DNP approach for the first demonstration of RDNMR in a 2DEG within the typical narrow-gap semiconductor InSb. The large effective $g$ factor ($g^*$, over 39 in magnitude) of the InSb 2DEG increases the ratio of Zeeman to cyclotron energies and thereby contributes to the spin-flip Landau level (LL) intersection around the filling factor (the number of occupied LLs) of $\nu = 2$ in tilted magnetic fields. The simplest pseudospin quantum Hall ferromagnet (QHF) with domain structures is formed therein, involving the lowest two LLs with only the orbital (n = 0,1) and spin ($\sigma = \uparrow,\downarrow$) indices. The QHF is characterized by a resistance spike within persistent longitudinal resistance $R_{xx}$ minima. DNP occurs when a large current flows through the QHF, resulting in the successful RDNMR detection of the high-nuclear-spin isotopes of both In and Sb. The present results are of fundamental interest for a better understanding of the simplest pseudospin QHF and DNP in low-dimensional semiconductor structures and are of practical relevance for quantum information processing.

The 2DEG used here was confined to a 30-nm-wide InSb quantum well with symmetrically doped Al$_{0.09}$In$_{0.91}$Sb barriers[13] and was processed in a Hall bar (170×40 μm). RDNMR and tilted-magnetic-field measurements were performed using a dilution refrigerator with an *in situ* rotator. A small RF field (~ μT, continuous wave) for NMR studies was generated by a single turn coil wound around the sample and a standard AC lock-in technique at 13.3 Hz was used to measure $R_{xx}$. Our 2DEG had an electron density of $n_s = 1.83 \times 10^{15}$ m$^{-2}$ and a low-temperature electron mobility of $\mu = 14.3$ m$^2$/Vs.

As a 2DEG sample rotates around a magnetic field $B_{tot}$ [inset, Fig. 1(b)], the perpendicular field component $B_\perp$ drives electrons to occupy LLs separated by the cyclotron energy, $\hbar\omega_c = \hbar eB_\perp/m^*$ (where $\hbar$ is the reduced Planck's constant and $m^*$ is the effective mass). $B_{tot}$ also couples the electron spins and splits each LL into two sublevels with the Zeeman energy, $\hbar\omega_s = g^*\mu_B B_{tot} = g^*\mu_B B_\perp/\cos\theta$ (where $\mu_B$ is the Bohr magneton and $\theta$ is the tilt angle). The LL energy gap results in the vanishing of $R_{xx}$ as a signature of the IQH effect. The ratio of $\hbar\omega_c$ and $\hbar\omega_s$ can be tuned by tilting $\theta$. For a constant $B_\perp$ (or $\hbar\omega_c$), $\hbar\omega_s$ increases linearly with increasing $1/\cos\theta$ [Fig. 1(a)], leading to a spin-flip LL intersection at $r\hbar\omega_s = \hbar\omega_c$ with integer coincidence ratios $r$.[14] The LL energy gap collapses at the intersection in the absence of electron-electron interactions, showing a resistance peak instead of an $R_{xx}$ minimum. However, richer physics is expected at the LL intersection in the presence of strong electron-electron interactions[15-17] (see also below). Because $g^*$ of the InSb 2DEG is extremely large,[18] the remarkable enhancement of $\hbar\omega_s$ places the LL degeneracy at easily accessible fields, a phenomenon that motivated the studies reported here.

The evolution of LLs was mapped by measuring the peak positions of $R_{xx}$ at different $\theta$ with a small current of $I_{sd} = 3$ nA [Fig. 1(b)]. Two types of LL intersections were prominent. The first was characterized by a coalescence of $R_{xx}$ minima (circle), as also depicted in Fig. 1(a), and the other by a resistance spike within persistent $R_{xx}$ minima (solid line). The resistance spike occurred at the LL intersection in the region of high $B_\perp$ and small $\nu$, where the exchange energy is expected to be large.[16] Electrons therein prefer to occupy aligned LLs in the ordered state to gain the exchange energy prior the single-particle energy cross. Because the intersecting LLs have different orbit and spin indices, the pseudospin quantum number is used to describe the phase transition from a paramagnetic state to a ferromagnetic state.[15] For instance, pseudospin-down and pseudospin-up at the $\nu = 2$ LL intersection correspond to the states of $(n, \sigma) = (0,\downarrow)$ and $(n, \sigma) = (1, \uparrow)$. In this case, the QHF has easy-axis (Ising) anisotropy with discrete preferred pseudospin orientations (also called quantum Hall Ising ferromagnet). Disorder or a finite temperature produces a domain wall separating domains with different pseudospin polarizations.[17] Charge transport through the domain wall results in the resistance spike.[17,19] The $\nu = 2$ QHF is the simplest pseudospin QHF involving only the lowest two LLs, providing an ideal basis for understanding the physics of the QHF. However, the previous studies of the $\nu = 2$ QHF have been limited to conventional magnetotransport measurements.[13,20]

The dependence of $R_{xx}$ on $B_\perp$ around $\nu = 2$ is shown in Fig. 2(a). The resistance spikes (arrow) were evident between $\theta = 58.7°$ and $\theta = 61.2°$ in the magnetotransport measurement. We will see later that the following high-sensitivity RDNMR approach can even probe the spike inside the LL peak. The resistance spike was found to depend on $I_{sd}$ (Figs. S1 and S2 of Ref. 21). At relatively large $I_{sd}$, $R_{xx}$ exhibited an exponential increase over time [inset, Fig. 2(b)]. As $R_{xx}$ became saturated, the RF was set to irradiate the 2DEG while sweeping the frequency $f$ through the Larmor resonance of each nucleus. Figure 2(b) shows $R_{xx}$ vs. $f$ for $^{115}$In. A dip with a resistance change $\Delta R_{xx}$ appeared at resonance frequency $f_{NMR} = 65.42$ MHz, representing an RDNMR signal of $^{115}$In. In a similar way, the RDNMR signals of $^{121}$Sb and $^{123}$Sb were also obtained (Fig. S3 of Ref. 21). We did not observe the RDNMR signal of $^{113}$In as expected from its low nuclear abundance (4.3%). The $f_{NMR}$ value gives the gyromagnetic ratio $\gamma = 2\pi f_{NMR}/B_{tot}$: $\gamma_{^{115}In} = 58.7 \mathrm{MHz/T}$, $\gamma_{^{121}Sb} = 64 \mathrm{MHz/T}$, and $\gamma_{^{123}Sb} = 34.7 \mathrm{MHz/T}$, in accordance with those in bulk InSb.[22]

The observed RDNMR in the InSb 2DEG relies on the contact hyperfine interaction.[23] Because the spin reversal energy of nuclei is one-thousandth of that of electrons, electron-nuclear flip-flop requires a significant energy reduction in electron spin-flip. The domains with different pseudospin polarizations are energetically degenerate in the $\nu = 2$ QHF, probably allowing for the flip-flop. A large current is applied to drive more electrons to move from one domain to the next with spin flip, simultaneously making the relevant nuclei flop. Because the nuclear polarization occurs only at or near the domain wall with respect to electron spin-flip, the distributions of the DNP sites are inhomogeneous. This inhomogeneity induces additional domain disorder via the Overhauser shift,[24] leading to an increase of $R_{xx}$ due to extra dissipation between extended LL states. In contrast, the depolarization reduces the domain disorder and results in a decrease of $R_{xx}$. Therefore, the RDNMR shows a dip in $R_{xx}$ [Fig. 2(b)]. Note that the RDNMR signals at most of the LL intersections around $\nu = 2$ show a dip in $R_{xx}$, but exhibit a hump in $R_{xx}$ at the LL intersections at $\theta = 53.2°$ (data not shown) and $\theta = 64.3°$ [Fig. 3(a)]. The spike after the DNP (called polarized spike) might shift relative to the original one (unpolarized spike) owing to the influence of the LL peak, with the result that $R_{xx}$ in one flank of the polarized spike is smaller than that of the unpolarized spike, thereby accounting for the hump feature. The current dependence of $|\Delta R_{xx}|/R_{xx}$ of $^{115}$In is shown in Fig. 2(c), in which $|\Delta R_{xx}|/R_{xx}$ appears at ~ 0.6 $\mu$A and increases with $I_{sd}$ but decreases at higher $I_{sd}$ as expected from the current-induced heating effect. This result is indicative of a current-dependent DNP.

The regions with detectable RDNMR around $\nu = 2$ were identified by dots in Fig. 2(a). Apparently, the DNP always occurred around the resistance spike, indicating the domain structure crucial for DNP.

More interestingly, the high-sensitivity RDNMR allowed us to probe the spike inside the LL peak and to determine the whole area of the $\nu = 2$ QHF between $\theta = 53.2°$ and $\theta = 64.3°$.

Further reducing the RF sweep rate enabled the observation of the quadrupole splittings (QSs) of $^{115}$In. A distinct QS spectrum was observed in the RDNMR signal taken from the spike inside the LL peak [Fig. 3(a)]. The fit in Fig. 3(a), as indicated by a solid line composed of nine nearly equally spaced Gaussian curves (dashed line), was in good agreement with the data. These nine curves are responsible for resonant transitions among ten quadrupole levels of $^{115}$In with equal splitting intervals,[25] as depicted in Fig. 3(b). The splitting ($\omega_q/2\pi \sim 57$ kHz) in Fig. 3(a) gives an estimate of the deformation tensor element $e_{zz}$ of $\sim 0.2\%$ in the case of uniaxial strain along the growth direction,[26] suggesting low strain at relevant DNP sites. This estimate also coincides with our expectation that the lattice mismatch in our InSb QW is small ($\sim 0.7\%$).

The developed RDNMR also enables the measurement of the nuclear spin relaxation time ($T_1$) in the simplest pseudospin QHF. The procedure for measuring $T_1$[11,27] is shown in the inset of Fig. 4. Note that $T_1$ is not determined by the current-pumped DNP.[28] $T_1$ was found to be about 120 s and almost temperature ($T$) independent (Fig. 4). This $T_1$ value was nearly the same as that of the QHF in a two-subband GaAs 2DEG, where collective spin excitations in the domain wall account for the short $T_1$.[27] Furthermore, the $T$-independent $T_1$ suggests that the nuclear spin relaxation is dominated by disordered spin-textured quasiparticles[9,11]. In future work, the $T_1$ measurement of an InSb 2DEG with gate-controlled electron density is expected to elucidate the properties of the low-energy spin excitations in the simplest pseudospin QHF.

We also performed the RDNMR measurement at other LL intersections. No RDNMR was observed at the LL intersections marked with circles in Fig. 1(b). This result further suggests the domain required for the DNP at the LL intersection. However, we found that not all LL intersections with this domain structure favor DNP. RDNMR was not observed at the spikes around $\nu = 3$ and $\nu = 4$ [Fig. 1(b)]. It is possible that the partially polarized ground states $|(0,\uparrow),(1,\uparrow),(0,\downarrow)\rangle$ and $|(0,\uparrow),(1,\uparrow),(2,\uparrow),(0,\downarrow)\rangle$ of the $\nu = 3$ and $\nu = 4$ QHFs, respectively, weaken electron spin flip, and accordingly smear out the DNP. Moreover, the DNP can be suppressed by the mobility of domains, which is expected to be different at different LL intersections.[29]

In conclusion, we have successfully demonstrated DNP and RDNMR in a single InSb 2DEG. Pseudospin domain structures have been shown to account for the DNP in the simplest pseudospin QHF characterized by a relatively short and temperature-independent $T_1$. The obtained QSs of [115]In are of practical importance for the coherent control of the ten nuclear-spin quantum levels and for the implementation of multiple NMR-based quantum bits ($\geq 2$). Our results advance the understanding of DNP in semiconductors and the domain dynamics in the QHF, and pave the way for NMR studies addressing low-dimensional structures made of narrow-gap semiconductors.

We acknowledge helpful discussions with K. Hashimoto, N. Shibata, S. Watanabe, and K. Muraki. H.W.L. thanks the Program for New Century Excellent Talents of the University in China.

[28] We confirmed that $T_1$ did not depend on DNP by the following experiment. As DNP became saturated [inset, Fig. 2(b)], the current was turned off over a certain time ($\tau$) and then switched on to measure $R_{xx}$ immediately. By repeating this process in different time periods, we obtained the plot of $R_{xx}$ vs. $\tau$, from which $T_1$ was determined. $T_1$ measured in this way was comparable to that in the inset of Fig. 4.

FIG. 1 (color online). (a) Diagram of the Landau levels in the 2DEG, showing the intersection of spin-down ($\sigma = \downarrow$) and spin-up ($\sigma = \uparrow$) levels with different orbital numbers ($n$) as $1/\cos\theta$ [where $\theta$ is the tilt angle, inset of (b)] increases, but keeping the perpendicular field $B_\perp$ (or cyclotron energy $\hbar\omega_c$) constant. The level intersection occurs at integer $r$ ($=\hbar\omega_s/\hbar\omega_c$, $\hbar\omega_s$ is the Zeeman energy). Integers inside the diagram denote the filling factor $\nu$. (b) Peak positions (dots) of $R_{xx}$ vs. $B_\perp$ (or $\nu$) and $1/\cos\theta$ (or $\theta$) at $I_{sd}$ = 3 nA and $T$ = 200 mK. The circle guides the LL intersection with a coalescence of $R_{xx}$ minima and the solid line marks the resistance spike of QHF (see text). In contrast to (a), the unevenly spaced LLs were caused by the change of $B_\perp$. The dashed lines indicate the LL intersections at the same integer $r$. As $B_\perp$ increases, $g^*$ is enhanced,[18] moving the LL intersection to small $\theta$ (or $1/\cos\theta$) for fixed $r$. Inset: Setup of our tilted-field experiments. $\theta$ was determined from the Hall resistance.

FIG. 2. (a) $R_{xx}$ vs. $B_\perp$ at several angles [raw data around $\nu = 2$ in Fig. 1(b)]. The curves are offset for clarity. The arrows point out the distinguishable resistance spikes in the magnetotransport measurement. We also superimposed the data (dots) obtained from the RDNMR measurement, which pointed out the DNP region around $\nu = 2$. The dot diameter denotes the ratio between the DNP-induced resistance change $\Delta R_{xx}$ [(b)] in terms of the absolute value ($|\Delta R_{xx}|$) and $R_{xx}$ of $^{115}$In. (b) RDNMR signal of $^{115}$In at an RF sweep rate of 60 kHz/min and $T$ = 100 mK. The data were obtained from the spike at $\theta$ = 59.5° in (a). The RDNMR measurement was initiated by sweeping $B_{tot}$ to 7 T ($B_\perp$ = 3.55 T) at $I_{sd}$ = 3 nA. Then, a large current of 1.4 $\mu$A was applied and time dependent $R_{xx}$ was obtained (inset). The RF field was set to illuminate the 2DEG after $R_{xx}$ became saturated. (c) $|\Delta R_{xx}|/R_{xx}$ in (b) vs. $I_{sd}$.

FIG. 3 (color online). (a) RDNMR signal of $^{115}$In at a slow RF sweep rate of 3 kHz/min and $T$ = 100 mK. The data was obtained from the spike inside the LL peak at $\theta$ = 64.3° in Fig. 2(a). The solid line is a fit with nine Gaussian curves (dashed line). (b) Energy diagram of the ten nuclear spin states |m⟩ of $^{115}$In ($I$ = 9/2) in the

presence of electric quadrupole coupling. $\omega_0$ and $\omega_q$ are the Zeeman and quadrupole frequencies, respectively.

FIG. 4 (color online). Temperature ($T$)-dependent $T_1$ of $^{115}$In. As $R_{xx}$ became saturated [inset, Fig. 2(b)], the RF frequency was tuned to $f_{NMR}$ = 65.42 MHz (on resonance). $R_{xx}$ decreased due to the nuclear depolarization and then reached a steady state (inset). After that, $f$ moved away from $f_{NMR}$ (off resonance) and $R_{xx}$ decayed to the original value. The time ($t$) dependent $R_{xx}$ at off-resonance followed $R_{xx} \propto e^{-t/T_1}$, from which $T_1$ was determined. Note that the $T_1$ measurement was independent of the current amplitude (Fig. S4 of Ref. 21).

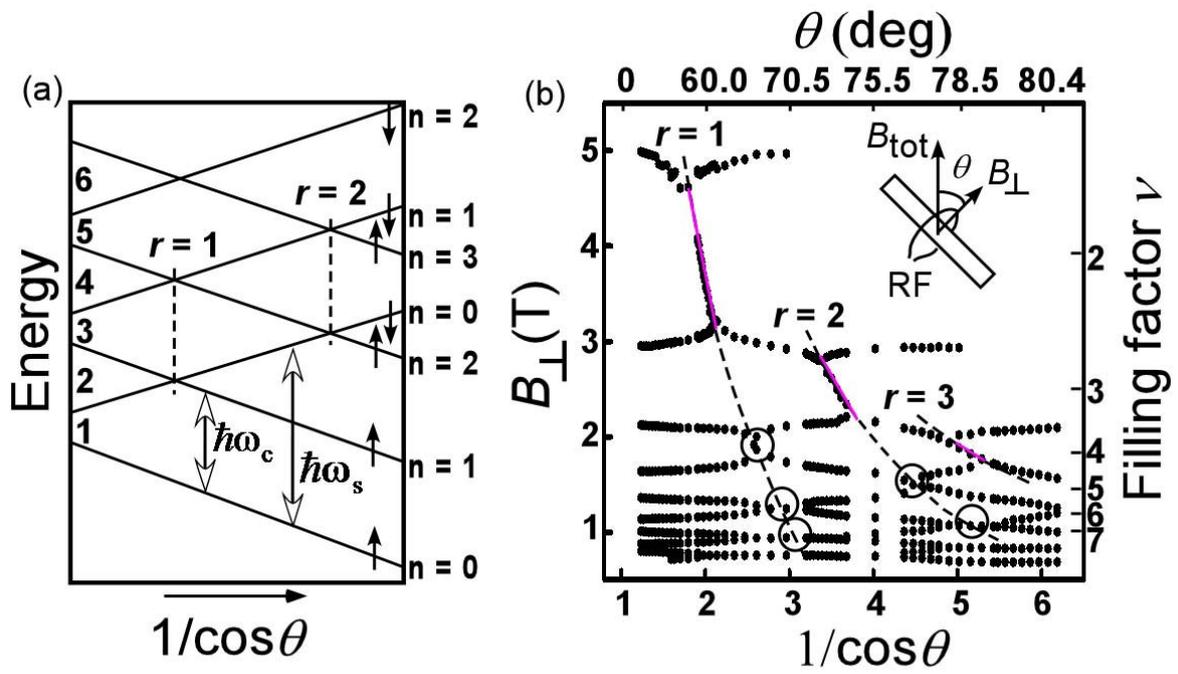

FIG.1

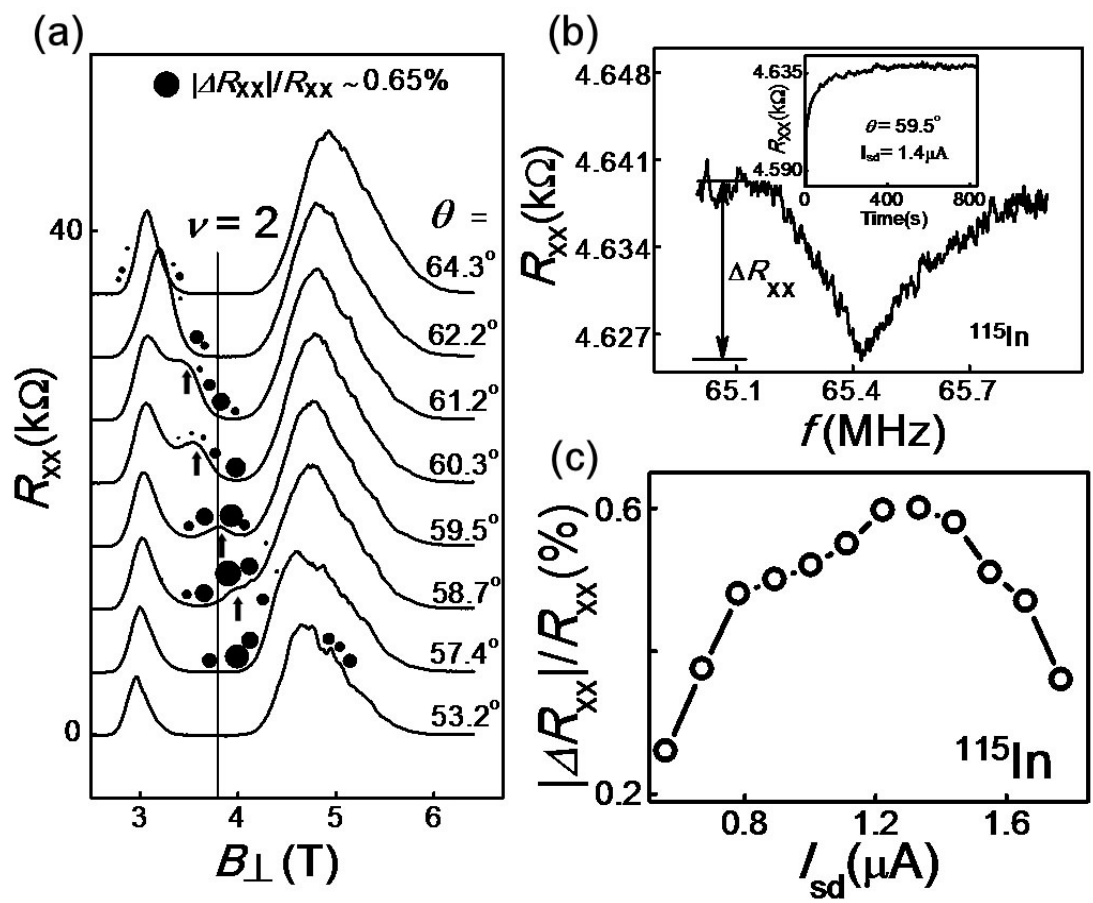

FIG. 2

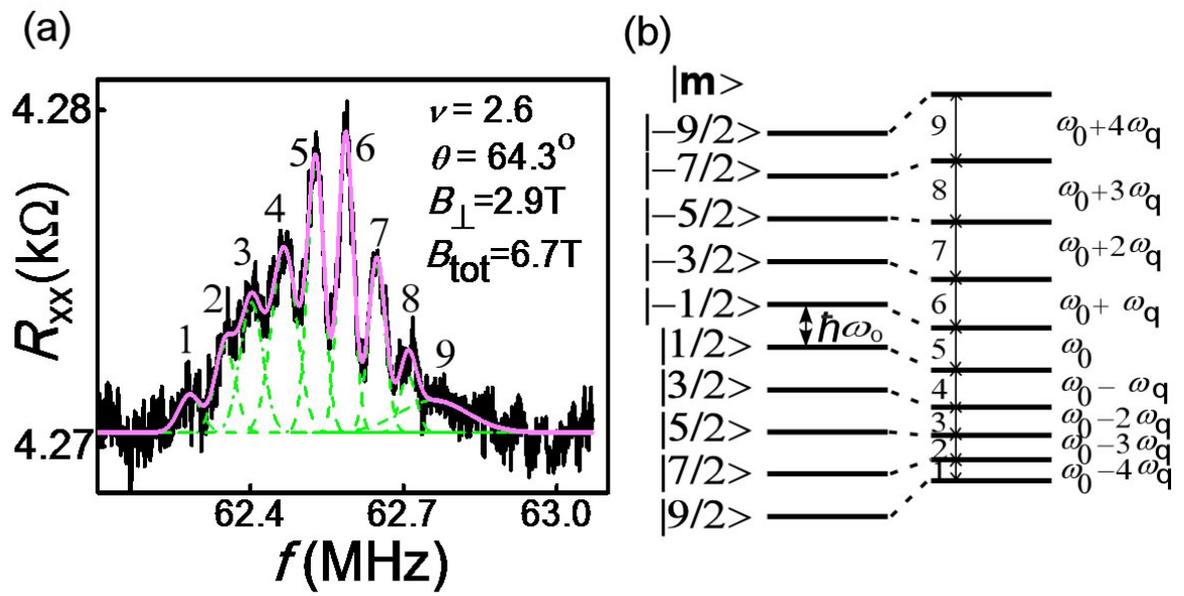

FIG. 3

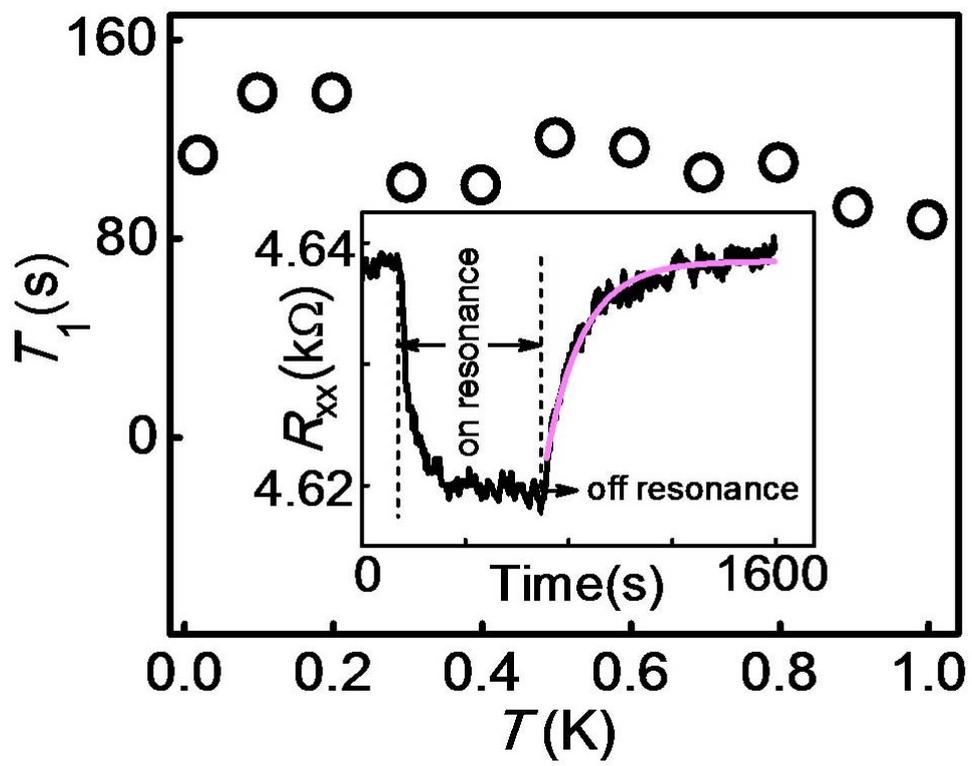

FIG. 4